\documentclass[preprint,12pt]{elsarticle}

\usepackage{amssymb}

\begin{document}

\begin{frontmatter}



\title{Generation of Projector Augmented-Wave atomic data: a 71 elements validated table in the XML format}


\author[CEA]{Fran\c{c}ois Jollet$^{\star}$ }
\author[CEA]{ Marc Torrent}
\author[WFU]{ Natalie Holzwarth}
\address[CEA]{ CEA, DAM, DIF F-91297 Arpajon, France}
\address[WFU]{Department of Physics, Wake Forest University, Winston-Salem, NC 27109 USA}

\date{\today}

\begin{abstract}
A Projector Augmented Wave (PAW) atomic data file is needed to be generated for each element, and plays in the PAW method the role of the pseudopotential file for norm-conserving (NC) or ultra-soft (US) plane wave calculations. In this paper, we present a review on how to obtain these data as well as results concerning their accuracy, their transferability and their efficiency for bulk solids. Following \cite{Lejaeghere}, we propose a new criterium to test PAW atomic data and we provide a new table written in a XML format potentially readable by every PAW electronic structure code.
\end{abstract}


\end{frontmatter}



\section{Introduction}
\label{}
In the context of Density Functional Theory (DFT) electronic structure calculations performed in the framework of the plane wave pseudopotential approach, the interaction between valence electrons and the ions formed by the atom nucleus and the core electrons is described by a potential called pseudopotential. This pseudopotential is an atomic quantity that is provided for each element. Many efforts have been done for thirty years to calculate them, improving their accuracy and their transferability to many systems. In 1979, Hamann, Schl\"uter and Chiang \cite{Hamann} introduced the notion of norm-conserving (NC) pseudopotentials and many people have proposed different ways to calculate them (see for instance \cite{Kleinman},\cite{Troullier},\cite{Goedecker}). At this stage, the pseudopotential files mainly contain pseudo wavefunctions and corresponding pseudopotentials for each relevant l quantum number. In order to decrease the number of plane wave sometimes needed for NC calculations, Vanderbilt introduced the concept of ultra-soft (US) pseudopotential, for which the norm-conserving condition has been relaxed \cite{VanderbiltUSPS}, whereas, Bl\"ochl~\cite{Blochl} proposed a new formalism called projector augmented-waves (PAW). Doing this, they introduced the notion of projectors that are atomic functions that must be given in the pseudopotential file. The PAW method is based on a linear transformation allowing to express the all-electron (AE) wavefunction of  an electronic state as a function of a pseudo-wavefunction (PS) developped on a plane wave basis, of AE and PS atomic partial waves and of associated projectors. This relation allows to theoretically well-found the PAW formalism, showing it is directly derived from an all-electron formalism thanks to a well-controlled approximation. Moreover, it can be shown that the US and NC formalisms can be derived from the PAW formalism\cite{KresseJoubert}. 

In the original PAW formalism proposed by Bl\"ochl~\cite{Blochl}, it is therefore necessary to provide for each element the AE and PS partial waves and the associated projectors. 
As it is  required to provide both atomic basis and projector functions as well
as    pseudopotential 
information, we prefer to speak about "PAW atomic data", rather than "PAW pseudopotentials".

The PAW method has been implemented in several codes (see for instance \cite{Blochl}, \cite{Holzwarth1}, \cite{KresseJoubert}, \cite{Mortensen},\cite{Abinit},\cite{Giannozzi}). The PAW data they are using are obtained on partial waves and projectors based  either on Bl\"ochl's \cite{Blochl} scheme or on US scheme \cite{Laasonen} (called here Vanderbilt's scheme). At the end, efficient PAW atomic data files based on a 
RRKJ pseudisation scheme \cite{RRKJ} 
are provided by the Vienna ab initio simulation package 
(\cite{Vasp}, \cite{KresseJoubert}, \cite{Furth}) 
the particularity of which is to include a compensation charge $\hat{n}$ in the calculation of the exchange and correlation potential. This large diversity of approaches, added to the wellknown difficulty to generate efficient and transferable pseudopotentials, makes the generation of PAW atomic data files tedious. Moreover, one needs sometimes special data for very specific problems: indeed, when one wants to study for instance materials under high pressure, it is necessary to have very small augmentation radii and semi-core electrons in the valence, and therefore, to generate non-standard atomic data. To do this, one is willing to compare the different approaches to choose the most suited to the studied case.

In this paper we present a review of the different schemes available to generate PAW atomic data and a new table we have generated for 71 elements, as well as results concerning its accuracy, its transferability and its efficiency for bulk solids.

The paper is organised as follows: In Sec 2, the general background and formalism of the different schemes to obtain wavefunctions and projectors are presented. Section 3 gives the methodology we have followed to generate a new PAW atomic data table whereas section 4 is devoted to the validation and efficiency of this table when compared with two other packages, thanks to the so-called delta factor \cite{Lejaeghere} and a modified delta factor that we will introduce.

\section{General background and formalism}
 
The details of the PAW method have already been given several times in the litterature (\cite{Blochl},\cite{Holzwarth},\cite{KresseJoubert}). We have adopted the notations of \cite{KresseJoubert} and refer also to a previous paper \cite{Torrent}. 

The PAW method is based on a linear transformation linking the AE (all electron) wavefunctions $\Psi_{\rm nk}$, with the n index corresponding to a summation over the bands and k indexing the k-points, to the PS(pseudo)one:	

\begin{equation}
\mid \Psi_{\rm nk}>= \mid \tilde \Psi_{\rm nk}>+\sum _{\rm i}(\mid \phi_{\rm i}>-
\mid \tilde \phi_{\rm i}>)<\tilde p_{\rm i}\mid\tilde\Psi_{\rm nk}>
\label{transformation}
\end{equation}

The index i stands for the atomic position $\vec R$, the angular momentum (l,m) and an additional index to label different partial waves for the same site and angular momentum. The AE $\phi_{\rm i}$ and PS $\tilde \phi_{\rm i}$ partial waves are atomic functions defined in a sphere and equal outside the augmentation region. Therefore, as in the NC (norm-conserving) scheme, $\Psi_{\rm n}= \tilde \Psi_{\rm n}$ outside a core radius $r_{\rm n}$. The projector functions $\tilde p_{\rm i}$ are dual to the partial waves:

\begin{equation}
<\tilde p_{\rm i}\mid\tilde\phi_{\rm j}>=\delta_{\rm ij}
\end{equation}

The wavefunctions $\tilde \Psi_{\rm nk}$ are solutions of the eigenvalue equation:

\begin{equation}
\tilde{H}\tilde \Psi_{\rm nk} =\epsilon_{\rm nk} O\tilde \Psi_{\rm nk}
\end{equation}

with:
\begin{equation} 
O=1+\sum_{\rm ij}\mid \tilde p_{\rm i}>(<\phi_{\rm i}\mid\phi_{\rm j}>)
-<\tilde \phi_{\rm i}\mid\tilde \phi_{\rm j}>)<\tilde p_{\rm j}\mid
\end{equation} 
\begin{equation}
\tilde{H}=-\frac{1}{2}\Delta + \tilde{v}_{\rm eff}+ \sum_{\rm ij}|\tilde{p}_{\rm i}>
D_{\rm ij}<\tilde{p}_{\rm j}| \\
\end{equation}
where $\tilde{v}_{\rm eff}=
v_{H}[\tilde{n}_{Zc}] + v_{H}[\tilde{n}+\hat{n}] + 
v_{xc}[\tilde{n}+\tilde{n}_{c}]$ is an effective local potential 
and $D_{\rm ij}$ is a non-local term. $v_{H}$ is the Hartree potential 
and $v_{xc}$ the exchange correlation potential.

In the above equations, the quantities  $\phi_{\rm i}, 
\tilde \phi_{\rm i},\tilde p_{\rm i},\hat{n},\tilde{n}_{\rm c}, 
v_{\rm H}[\tilde{n}_{\rm Zc}] $ are atomic quantities that 
must be provided in the PAW atomic data file for each element. 
$\tilde n_{\rm c}$ is the pseudization of core electron charge density 
$n_{\rm c}$. The contribution $v_{H}[\tilde{n}_{Zc}]$ represents
pseudization of Coulombic contributions of the nuclear charge Z and 
the pseudo core density.

\subsection{All electron atomic calculations}

As in other approaches, the first step to generate PAW atomic data is to perform an AE calculation for a given atomic configuration. All wavefunctions (and projectors) are written: 
\begin{equation}
\phi_{\rm i}(\vec r)=\frac{\phi_{\rm n_i \rm l_i}(r)}{r}S_{\rm l_i m_i}(\hat r) 
\end{equation}
with $ S_{\rm lm}(\hat r)$ the real spherical harmonics. It is assumed that the total electron density can be partitioned into a core electron density $n_{\rm c}(r)$ corresponding to $Q$ electrons and a valence electron density. The core density is assumed to be "frozen" in the same form in the solid as it is in the atom. The radial atomic eigen wavefunctions $\phi_{\rm i}$  are solution of:
\begin{equation}
[T+V_{\rm AE}(r)] \phi_{\rm i}(r)=\epsilon_{\rm i} \phi_{\rm i}(r)
\label{hamiltonian1}
\end{equation}
where T is the kinetic energy operator and $V_{\rm AE}$ the AE potential.

In practice, the $ {\phi_{\rm i}}$  are solutions of a non-relativistic Schrodinger equation or for heavier elements of a (scalar-)relativistic equation as developed by Koelling and Harmon \cite{Koelling}.
The transformation (\ref{transformation}) implicitely supposes that the partial wave basis is complete.
As usual, a compromise must occur between accuracy and efficiency to select the number of partial waves included in the partial wave basis. Most of the time, a reasonable choice is to take two partial waves per angular momentum for ground state calculations. 
This implies to select a set of energies $\epsilon_{\rm i}$  (i indexes the different partial waves for momentum ${\rm l}$) and of cutoff radii. Equation (\ref{hamiltonian1}) is then inverted, which builds a set of AE partial wave ${\phi_{\rm i}}$ (one usually takes an eigen state among the two partial waves per angular momentum, but it is not compulsory). 

\subsection{Potential pseudization}

The next step into consideration is to build a pseudopotential function $V_{\rm PS}$ that will be used to obtain the smooth basis function and whose unscreened version will appear as a local pseudopotential. It is built so that it matches smoothly to $V_{\rm AE}$ at a cutoff radius ${r_{\rm loc}}$.

Several ways have been proposed in the litterature to obtain $V_{\rm PS}$ ( \cite{Blochl}, \cite{Holzwarth1}, \cite{KresseJoubert}).
The first one is to use a Troullier-Martins norm-conserving scheme \cite{Troullier}. We choose first a reference energy $E_{\rm loc}$ and a reference angula momentum $\rm l_{\rm loc}$. A PS wavefunction is chosen to have the form 
\begin{equation}
\phi_{\rm PS}(r)= r^{\rm l_{\rm loc}+1}exp(p(r))
\end{equation}
for $r<r_{\rm loc}$, where p is an even $12^{\rm th}$ polynomial. $\phi_{\rm PS}$ matches continuously to AE wavefunction $\phi_{\rm loc}$ at $r=r_{\rm loc}$. Then, $V_{\rm PS}$ is deduced by inverting the wave equation at $\rm l=\rm l_{\rm loc}$ and $E=E_{\rm loc}$.

The second one is to use an ultrasoft scheme without norm conservation constraint. A PS wavefunction is chosen to have the form 
\begin{equation}
\phi_{\rm PS}(r)= r^{\rm l_{\rm loc}+1}\sum_{\rm m=0}^{\rm 3} C_{\rm m} r^{\rm 2m}
\end{equation}
for $r<r_{\rm loc}$. Then, $V_{\rm PS}$ is deduced by inverting the wave equation at $\rm l=\rm l_{\rm loc}$ and $E=E_{\rm loc}$.

A third possible scheme \cite{KresseJoubert} is to simply derive $V_{\rm PS}$ from $V_{\rm AE}$ using a zero-order Bessel function:
\begin{equation}
V_{\rm PS}(r)=\alpha \frac{sin(qr)}{r}
\end{equation}
for $r<r_{\rm loc}$.
For s or p elements, the norm-conserving scheme with $\rm{l}_{\rm loc}=\rm l+1$ is often a good choice as it has good scattering properties. For d or f elements, this may lead to ghost states and a direct pseudization of $V_{\rm AE}$ is often preferable. 

\subsection{The pseudization of the wavefunctions}
\subsubsection{The Vanderbilt scheme}

This scheme has been introduced by Vanderbilt in the formulation 
of US pseudopotentials  \cite{Laasonen} in which projectors are to be 
generated. 
A set of pseudized functions ${\tilde \phi_{\rm n_i l_i}}$ is first obtained 
so that each $\tilde \phi_{\rm n_i l_i}$ matches smoothly to 
$\phi_{\rm n_i l_i}$ at radius $r_{\rm n_i l_i}$. 
As for local potential, this pseudization can be performed in several ways. The form of the pseudized function can be either a polynomial:
\begin{equation}
\tilde \phi_{\rm n_i l_i}(r)= r^{\rm l_{i}+1}\sum_{\rm m=0}^{\rm p} C_{\rm m} r^{\rm 2m}
\end{equation}
or, following a RRKJ scheme \cite{RRKJ}, 
a sum of spherical Bessel functions:
\begin{equation}
\tilde \phi_{\rm n_i l_i}(r)= \sum_{\rm i=1}^{\rm p} r \alpha_{\rm i} j_{\rm l_{i}}(q_{\rm i}r)
\end{equation}
A slightly modified RRKJ scheme (modRRKJ) was recently introduced into the
ATOMPAW code \cite{Holzwarth2},  designed to ensure that the corresponding projector
functions have continuous first derivatives
 and  to control the desired number of nodes in the pseudo basis
functions \cite{web4}. The modRRKJ scheme needs further testing and has not
been used in the current study.\

For each smooth basis function, a localized auxiliary function can be formed:
\begin{equation}
\mid \chi_{\rm i}>= (\epsilon_{\rm i} -T-V_{\rm PS})\mid \tilde \phi_{\rm i}>
\end{equation}
which by design vanishes for $r>r_{\rm c}= Sup(r_{\rm i},r_{\rm loc})$.
The projector functions are then formed from a linear combination of these auxiliary functions of the same angular momentum:
\begin{equation}
\tilde p_{\rm j}(r)= \sum_{\rm i} \chi_{\rm i}(B^{\rm -1})_{\rm i,j}
\end{equation}
where the elements of the matrix B are given by:
\begin{equation}
B_{\rm i,j}= < \tilde \phi_{\rm i}\mid\chi_{\rm j}>
\end{equation}
As shown by Vanderbilt \cite{VanderbiltUSPS}, this construction ensures that
\begin{equation}
< \tilde \phi_{\rm i}\mid\tilde p_{\rm j}>=\delta_{\rm ij}
\end{equation}
and that the smooth function $\tilde \phi_{\rm i}(r)$ is an 
eigen function of the atomic PAW Hamiltonian.

\subsubsection{The Bl\"ochl scheme}

In Bl\"ochl's pseudofunction construction scheme \cite{Blochl}, the projector functions are built with the help of a shape function $k(r)$ that vanishes outside the augmentation region. It can be for example:

\begin{equation}
k(r)=\left\{
\begin{array}{l}
[\frac{sin(\pi r/r_{\rm c})}{(\pi r/r_{\rm c})}]^{\rm 2} \;\;\;  \mbox{for $r<r_{\rm c}$} \nonumber\\
0                    \;\;\;\;\;\;\;\;\;\;\;\;\;\;\;\;\;\;\;  \mbox{for $r>=r_{\rm c}$}
\end{array}
\right.
\end{equation}
The pseudo-basis functions ${\tilde \phi_{\rm i}(r)}$ are found by solving a self-consistent Schr\"odinger-like equation:
\begin{equation}
(T+V_{\rm PS}-\epsilon_{\rm i})\mid \tilde \phi_{\rm i}>=C_{\rm i}k(r)\mid \tilde \phi_{\rm i}>
\end{equation}
with $C_{\rm i}$ adjusted so that the logarithmic derivatives of 
${\phi_{\rm i}(r_{\rm c})}$ and ${\tilde \phi_{\rm i}(r_{\rm c})}$ are equal. The corresponding projector functions are then formed according to:
\begin{equation}
\tilde p_{\rm i}(r)= \frac{k(r)\tilde \phi_{\rm i}(r)}{< \tilde \phi_{\rm i} \mid k \mid \tilde \phi_{\rm i}>}
\end{equation}
so that $<\tilde p_{\rm i}\mid\tilde \phi_{\rm i}>=1$.
The final basis and projector functions $\phi_{\rm i}$, $\tilde{\phi_{\rm i}}$ and $\tilde{p_{\rm i}}$ are then obtained from the previous ones by a Gram-Schmidt orthogonalisation procedure.

\subsection{Core densities and unscrened potential}

The last quantities that must be provided in the PAW atomic data file are $n_{\rm c}$, $\tilde{n}_{\rm c}$ and $v_{\rm H}[\tilde{n}_{\rm Zc}] $. $n_{\rm c}$ is directely obtained from the AE calculation. It is then pseudized thanks to a polynomial scheme to obtain $\tilde{n}_{\rm c}$. 
The case of $v_{\rm H}[\tilde{n}_{\rm Zc}] $ is more delicate. It is obtained by unscreening $V_{\rm PS}$. Two schemes are reported in the litterature:

The first one is due to Bl\"ochl \cite{Blochl}. It implies:
\begin{equation}
\bar{v}= V_{\rm PS}-v_{\rm H}[\tilde{n}+\tilde{n}_{\rm c}+\hat{n}_{\rm B}]-v_{\rm xc}[\tilde{n}+\tilde{n}_{\rm c}]
\end{equation}
where
$\hat{n}_{\rm B}$ is a compensation charge that is added to the soft charge density $\tilde{n}$ to reproduce the correct multipole moment of the AE charge density.
We must therefore know $\hat{n}_{\rm B}$ in the atomic case to unscreen
 the potential in the same way it will be screened in periodic calculations. 
Following here the definition of $\tilde{n}$ and $\hat{n}$ given 
in \cite{Blochl}, we have for the atomic case:
\begin{equation}
\hat{n}_{\rm B}(r)=g_{\rm 0}(r)
\int_0^{\rm r_c}[n(r')-\tilde{n}(r')+n_{\rm c}(r')-\tilde{n}_{\rm c}(r')+
n_{\rm Z}(r')] dr'
\end{equation}
where $n_{\rm Z}$ is the nucleus charge density.\\
Several choices are possible for the shape of the compensation charge
function $g_{\rm l}(r)$ (for the atomic case, $\rm l=0$):
\begin{equation}
g_{\rm l}(r)=Nr^{\rm l}k(r)   \;\;\;  \mbox{with} \nonumber\;\;\; k(r)=[\frac{sin(\pi r/r_{\rm c})}{(\pi r/r_{\rm c})}]^{\rm 2}
\end{equation}
\begin{equation}
g_{\rm l}(r)=Nr^{\rm l}k(r)   \;\;\;  \mbox{with} \nonumber\;\;\; k(r)=exp(-(r/d)^{\rm 2})
\end{equation}
or 
\begin{equation}
g_{\rm l}(r)=\sum_{\rm i=1}^{\rm 2}  \alpha_{\rm i} j_{\rm l}(q_{\rm i}r)
\end{equation}
With these definitions, $\bar{v}$ is a potential localized in the PAW sphere. \\
The second one is due to Kresse and Joubert \cite{KresseJoubert}:
\begin{equation}
v_{\rm H}[\tilde{n}_{\rm Zc}]= V_{\rm PS}-v_{\rm H}[\tilde{n}+\hat{n}_{\rm K}]-v_{\rm xc}[\tilde{n}+\hat{n}_{\rm K}+\tilde{n}_{\rm c}]
\end{equation}

with 
\begin{equation}
\hat{n}_{\rm K}(r)=g_{\rm 0}(r) \int_0^{\rm r_c}[n(r')-\tilde{n}(r')] dr'
\end{equation}

The difference with the Bl\"ochl's formulation is that $\hat{n}$ has not the same definition in the Bl\"ochl's scheme ($\hat{n}_{\rm B}$) than in the Kresse-Joubert's one ($\hat{n}_{\rm K}$) and that $\hat{n}$ has been also included in the $v_{xc}$ term. It can be shown that the two formalisms are equivalent if we put:
\begin{equation}
v_{\rm H}[\tilde{n}_{\rm Zc}]= \bar{v}+ v_{\rm H}[\tilde{n}_{\rm c}+
\left(g_{\rm 0}/4\pi \right) (Q_{\rm core}-Z)]
\end{equation}
in the Bl\"ochl's formulation, with $Q_{\rm core}=\int_0^{\rm r_c}[n_{\rm c}-\tilde{n}_{\rm c}]dr$ and $Z=\int_0^{\rm r_c}[n_{\rm Z}]dr$, and if $\hat{n}_{K}$ is not included in the $v_{xc}$ term. With these definitions, $v_{\rm H}[\tilde{n}_{\rm Zc}]$ behaves like $(Q-Z)/r$, where $Q=\int_0^{\infty}[n_{\rm c}]dr$, for large r values.
There is no reason to include $\hat{n}$ in the $v_{xc}$ term but for a practical reason: manipulate in the code the only quantity $\tilde{n}+\hat{n}$ rather than the quantities $\tilde{n}$ and $\hat{n}$ separately. In practise, this may lead however to different physical results under certain conditions \cite{Torrent1}.

\section{PAW atomic data generation}
 
For each element, PAW atomic data generation follows the same steps as norm-conserving or US pseudopotentials. The first thing is to choose the exchange-correlation functional (LDA-PW or GGA-PBE for instance) and to select a scalar-relativistic wave equation. \\

\subsection{Electronic configuration}

The core and valence electrons must then be selected: in a first approach, we select only electrons from outer shells. But, if particular thermodynamical conditions are to be simulated, it is generally needed to include semi-core states in the set of valence electrons. Semi-core states are generally needed with transition metal and rare-earth materials. There are also some cases where physical conditions do not indicate a need for semi-core states, but the use of semi-core states is needed to avoid the appearance of the dreaded ghost states. Note that all wave functions designated as valence electrons will be used in the partial-wave basis. An excited configuration may be useful if the PAW dataset is intended for use in a context where the material is charged (such as oxides). However, in our experience, the results are not highly dependent on the chosen electronic configuration.

\subsection{Choice of the grid and of pseudization radii}
It is recommended to use a logarithmic grid for the AE  atomic calculation in order to well describe the region close to the nucleus. Indeed, it is very important, for instance, that the integral of the core density gives the number of core electrons with a very good accuracy. \\
Pseudization cutoff radii for each valence orbital, for local potential, for core density and for shape function are then chosen. It is well-known that the choice of cutoff radii is crucial for the efficiency of the PAW atomic data and that a compromise must be found between a too low radius, that in principle is better for accuracy but that requires a large plane wave energy cutoff, and a too large one, that allows to have a smaller energy cutoff, but often produces worse physical results, especially in case where PAW spheres are overlaping. Another tool to reduce the energy cutoff is to use Fourier filtering, as proposed for instance in  \cite{King}, but we have not used it in this study.\\

\subsection{Generation of the partial wave basis}

If necessary, supplementary partial waves are added, associated to a reference energy. We have generally chosen to have 2 partial-waves per angular momentum in the basis (this choice is not necessarily optimal but this is the most common one). When a good description of the conduction band is needed (excited states calculations for instance), 3 partial-waves per angular momentum may be needed. \\

The pseudization scheme for wavefunctions is then chosen (Bl\"ochl, Vanderbilt polynomial or RRKJ). We have noticed that Bl\"ochl scheme for projectors can produce very accurate datasets but sometimes with a low efficiency (in the sense that they may need a large number of plane waves to converge the DFT calculation). To increase performance, the Vanderbilt scheme is often better and the gain can be noticeable. But, most of the time, the best choice (for performance) would be the RRKJ scheme.\\

Running a PAW atomic calculation already gives good informations concerning the accuracy of PAW atomic data: it is indeed recommended that the partial-waves, pseudized partial-waves and projectors have an amplitude of the same order to avoid numerical instability and to promote a good transferability. If it is not the case, several options are possible:\\
- to change the matching radius for this partial-wave; but this is not always possible (spheres cannot have a large overlap in the solid)\\
- to change the pseudopotential scheme (see later).\\
- to move the reference energies away from each other if there are two (or more) partial waves for the considered ${\rm l}$ angular momentum, including additional partial waves (unbound states). This generally reduce the magnitude of projectors, but a too big difference between energies can lead to wrong logarithmic derivatives (see following paragraph).\\
To have accurate representation
 properties, PAW atomic data must lead to logarithmic derivatives of the eigen functions of the PAW atomic hamiltonian that are superimposed as much as possible to these of the exact atomic problem (for each ${\rm l}$-quantum number included in the partial wave basis). By construction, they are superimposed at the two energies corresponding to the two ${\rm l}$ partial-waves. If the superimposition is not good enough, the reference energy for the second ${\rm l}$ partial-wave should be changed.
A discontinuity in the logarithmic derivative curve often appears at 0$<$=$E_0$$<$=4 Rydberg. A reasonable choice is to choose the two reference energies so that $E_0$ is in between (if possible, i.e. if one the two partial-waves correspond to an unbound state).
Too close reference energies produce "hard" projector functions. But moving reference energies away from each other can damage accuracy of logarithmic derivatives.
Another possible problem is the presence of a discontinuity in the PAW logarithmic derivative curve at an energy where the exact logarithmic derivative is continuous. Most of the time, this shows the presence of a "ghost state". To avoid this, it is possible to change the value of reference energies; this sometimes can make the ghost state disappear. If not, we have to pay attention to the pseudization of the local potential.

\subsection{Pseudization of the local potential}
As already seen above, the pseudization scheme for local potential can be chosen among Troullier-Martins, ultrasoft of bessel schemes.  Norm-conserving pseudopotentials are sometimes so deep (attractive near r=0) that they produce "ghost states". A first solution is to change the l quantum number used to generate the norm-conserving pseudopotential. But this is generally not sufficient. Changing the pseudopotential scheme is (in most cases) the only efficient cure.
Selecting a simple bessel pseudopotential can solve the problem. But, in that case, one has to noticeably decrease the matching radius ${r_{\rm loc}}$ if one wants to keep reasonable physical results. Loosing to much norm for the wave function associated to the pseudopotential can have dramatic effects on the results.
Selecting a value of ${r_{\rm loc}}$ between 0.6*rpaw and 0.8*rpaw is a good choice; but the best way to adjust ${r_{\rm loc}}$ value is to have a look at the values of the valence energy obtained from a PAW atomic calculation compared to a reference atomic all-electron (AE) calculation. They have to be as equal as possible and are sensitive to the choice of ${r_{\rm loc}}$.\\

\section{PAW atomic data validation and efficiency}

\subsection{PAW atomic data validation}

\begin{center}
\begin{figure}[h]
{\resizebox{5.0cm}{!}
{\rotatebox{0}{\includegraphics{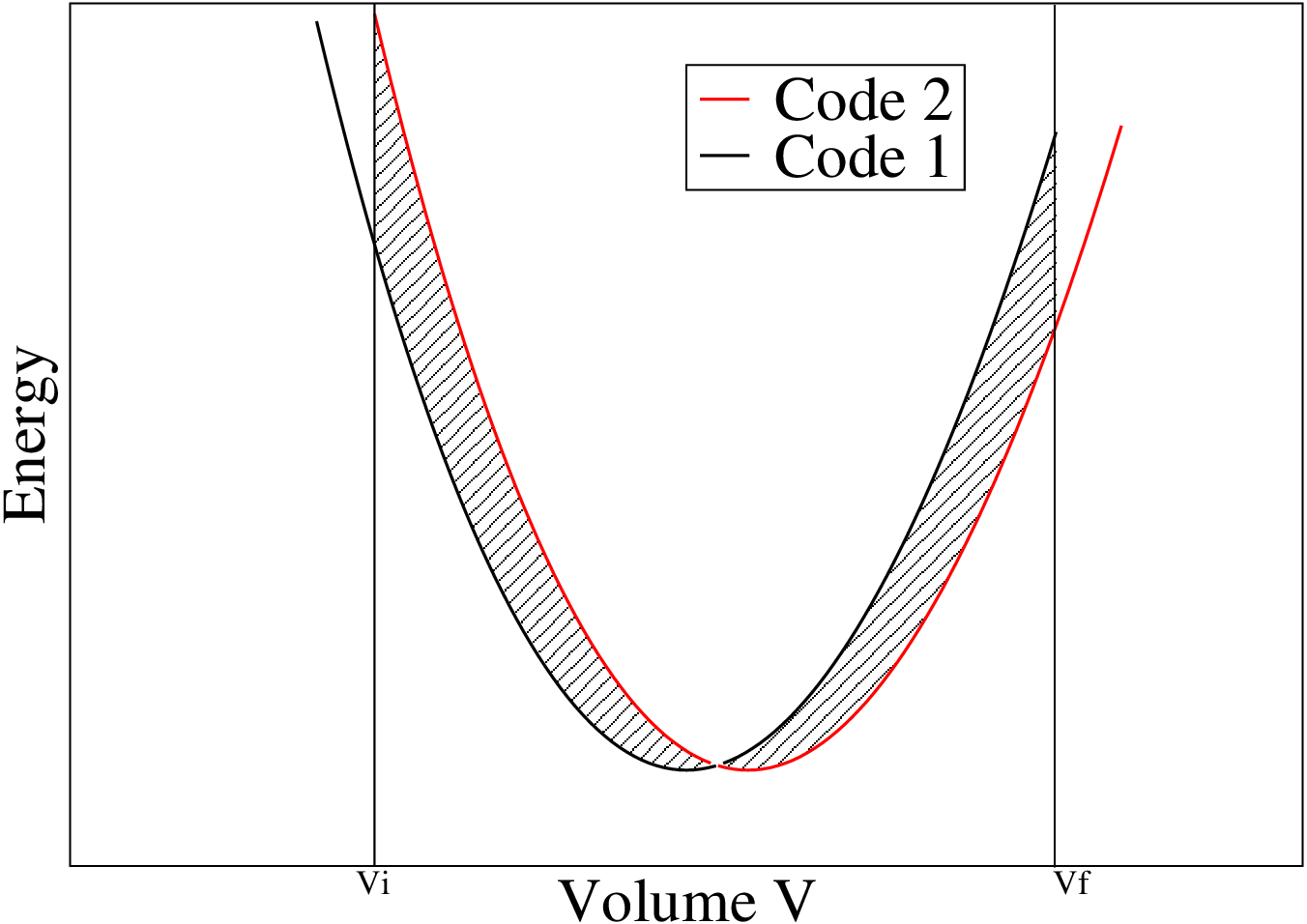}}}} \\
\caption{Comparison of the EOS obtained by two codes}
\label{fig1}
\end{figure}
\end{center}

 The PAW method is an all-electron method that uses auxiliary functions (plane waves for instance) as working functions. If the basis of partial waves is complete, PAW results must be in agreement with reference all electron calculations. The question of the measurement of the agreement of PAW and AE calculation is still a subject of debate. Most of the time, a solid state calculation is performed and the equilibrium volumes and bulk moduli are compared. This can be done on several environments like metals or oxides as for instance has been made recently for GBRV potentials \cite{Garrity}. To have a more flexible tool that allows comparisons between codes and between PAW atomic data tables, Lejaeghere et al. have recently introduced a new measure, named $\Delta$, of the agreement between two codes for a given structure \cite{Lejaeghere}. It is defined as the difference between the two equations of state, obtained by the two codes (see shaded region on fig.\ref{fig1}): 

\begin{equation}
\Delta=\sqrt{\frac{\int{\Delta{E^{\rm 2}}(V) dV}}{\Delta{V}}}
\end{equation}
\\
In this equation, $\int{\Delta{E^{\rm 2}}(V) dV}=\int_{V_{i}}^{V_{f}}{(E_{code1}(V)-E_{code2}(V))^{\rm 2} dV}$ and $\Delta{V}=V_{f}-V_{i}$ (see fig.\ref{fig1}).

This leads to a value of $\Delta$ for each element, from which a mean value on the whole table can be extracted. The results obtained for several codes and several pseudopotential tables are given on the web site of the authors \cite{web1}. It is also possible to download the delta calculation package that provides the results of solid state calculations for 71 elements with the AE code Wien2k (the equilibrium volume $V_{AE}$, the bulk modulus $B_{AE}$ and its derivative $B'_{AE}$), as well as the associated crystallographic data.\\
\\

We have therefore used the delta calculation package to validate our code and our new atomic data against the Wien2k code. The electronic structure calculations have been performed thanks to the ABINIT code \cite{Abinit}. For this we have used the recommended values \cite{Lejaeghere} for the k-point sampling (6750/N k-points in the Brillouin zone for a N-atom cell). A Fermi-Dirac broadenning of 0.002 Ha has been used. As indicated in \cite{Lejaeghere}, we have used the cystallographic data (CIF's files) provided with the delta calculation package. The Equation of State (EOS) of each element has been adjusted to a Birch-Murnaghan one thanks to seven calculations at seven different volumes, ranging from 0.94 to 1.06 $V_{\rm S}$, where $V_{\rm S}$ is the equilibrium volume deduced from the CIF's file, without geometry optimisation to be exactly in the same conditions as the Wien2k calculations.\\

The first step has been to validate the ABINIT code against the Delta calculation process. For this we have used the PAW atomic data PAW 0.9 provided on the GPAW web site \cite{web2}. These data being provided in a XML format, we have first coded the reading of XML files in ABINIT. Then, we have coded the radial grid used by the PAW 0.9 package: $r(i)=\frac{a*i}{n-i}$ with n, the number of points of the radial grid. At the end, we performed the ABINIT calculation for the 68 elements of the PAW 0.9 table. We finally obtained a mean value of $\Delta=1.6 meV$ for a cutoff energy of 20 Ha and 40 Ha. This value is very close to the value obtained with the GPAW code with the same PAW 0.9 package (1.8 meV) \cite{web1}, which validates the accuracy of the ABINIT code (the same as equivalent codes) and the Delta calculation process \\

\subsection{PAW atomic data table generation}

The second step has been to generate a new PAW atomic data table following the methodology described above in section 2 and 3.
We have generated a table of 71 PAW atomic data corresponding to elements ranging from H to Rn, without At and lanthanides (except Lu). All the PAW data have been obtained thanks to the ATOMPAW generator (v3.1.0.2) \cite{Holzwarth2}, starting either from existing input files already provided on the ABINIT web site \cite{web3}, or from new input files. All the schemes presented hereabove in section 2 are implemented in ATOMPAW, which makes it a very flexible tool.\\
For which concerns the choice of the electronic configuration, from H to Be, all the electrons are in valence electrons. For columns IA, IIA, IIIB to VIIIB of the periodic table of elements, semi-core states s and p are included in the valence. For colums IB and IIB, only s and d electrons are taken as valence electrons. For colums IIIA to VIIIA, only s and p electrons are taken as valence electrons. For Lu, the f electrons are also included in the valence electrons. For Pt, only s and d electrons are taken as valence electrons. \\
A logarithmic grid of the form r=a*(exp(d*i)-1) has been taken. At most 2 partial waves per angular momentum have been chosen. The cutoff radii have been chosen so that it leads to a standard cutoff energy of 20 Ha for plane waves in the ABINIT code. All the calculations have been performed with the GGA-PBE exchange and correlation functional in a scalar-relativistic framework. The crystallographic structures used for each element is described in \cite{Lejaeghere}.
\begin{center}
\begin{figure}[h]
{\resizebox{13.0cm}{!}
{\rotatebox{0}{\includegraphics{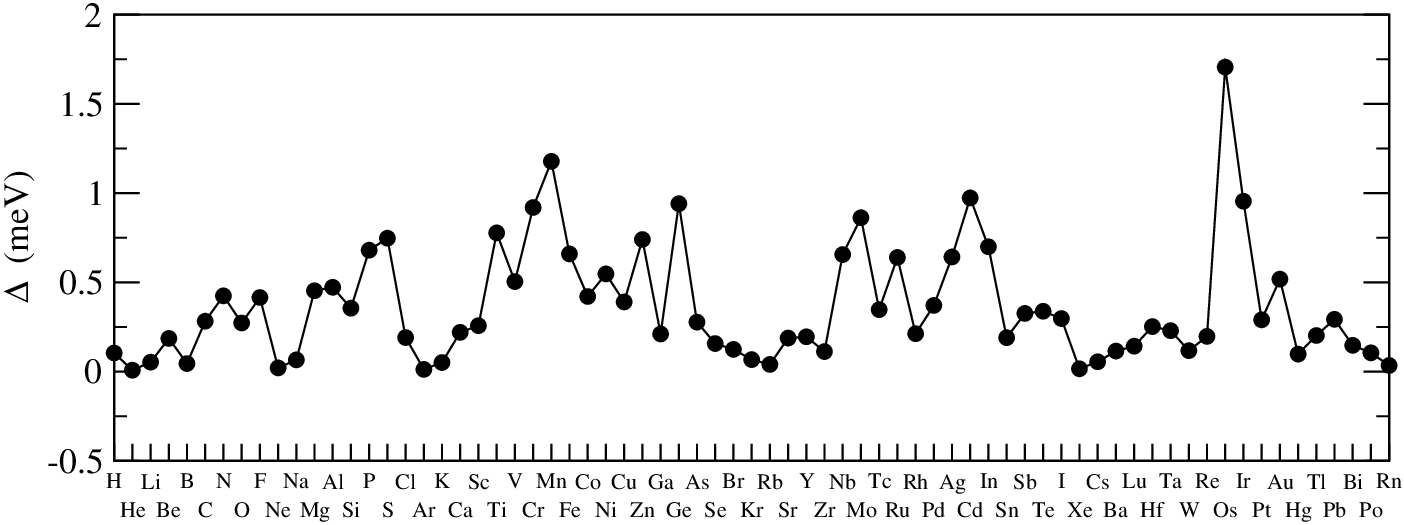}}}} \\
\caption{$\Delta$ value as a function of element for a 20 Ha energy cut-off for the JTH table}
\label{fig2}
\end{figure}
\end{center}

We obtain a mean value of $\Delta=0.4 meV$ for our whole table (named JTH table). The details for each element is given in fig.\ref{fig2}. The use of the $\Delta$ factor is very convenient: it allows to have a global measure of the accuracy of atomic data for each element, as well as a mean value that caracterizes a whole atomic dataset. The value of 0.4 meV is very good compared to values already published with other codes or other PAW atomic data packages for which $\Delta$ is the range 1.6-1.8 meV \cite{web1}\cite{Lejaeghere}.\\
However, one must be aware of some drawbacks using the $\Delta$ factor:\\
- The $\Delta$ value is by construction very dependent from AE calculations that are used for comparison. This means that we must be very confident in the AE results, which are also difficult and long to obtain. It is certainly necessary that the AE community agree on the tuning of the AE codes so that the values of $V_0$ (the equilibrium volume), B (the bulk modulus) and B' (the derivative of the bulk modulus) are well established for each element in the studied crystallographic structures. It would also be nice to add lanthanides and actinides in the AE references.\\
- The $\Delta$ factor is based on calculations on pure elements in their ground state crystallographic structures. It is also interesting to have comparison with AE calculations in compounds like oxides, as has been already done on the ATOMPAW web site \cite{web4} and by Garrity et al. \cite{Garrity}.\\
- In view of high-throughput calculations, it is extremely important to have efficient PAW atomic data. The $\Delta$ factor must therefore be given with an energy cut-off, and its convergency with the energy cut-off must be given.\\
- What is the accuracy of the $\Delta$ factor? What does it means to have $\Delta$=0.2 meV rather than 0.5 meV?\\
- We have noticed that for some elements, the $\Delta$ factor is very sensitive to the values $V_0$, B and B' obtained as a result of the tuning of the input parameters of the atomic data generator. This is not the case at all for other elements.\\
We therefore propose in the next section an alternative definition of a $\Delta$ factor to account for some of these drawbacks.

\subsection{PAW atomic data efficiency: towards a new $\Delta$ factor definition}

\begin{center}
\begin{figure}[h]
{\resizebox{5.0cm}{!}
{\rotatebox{0}{\includegraphics{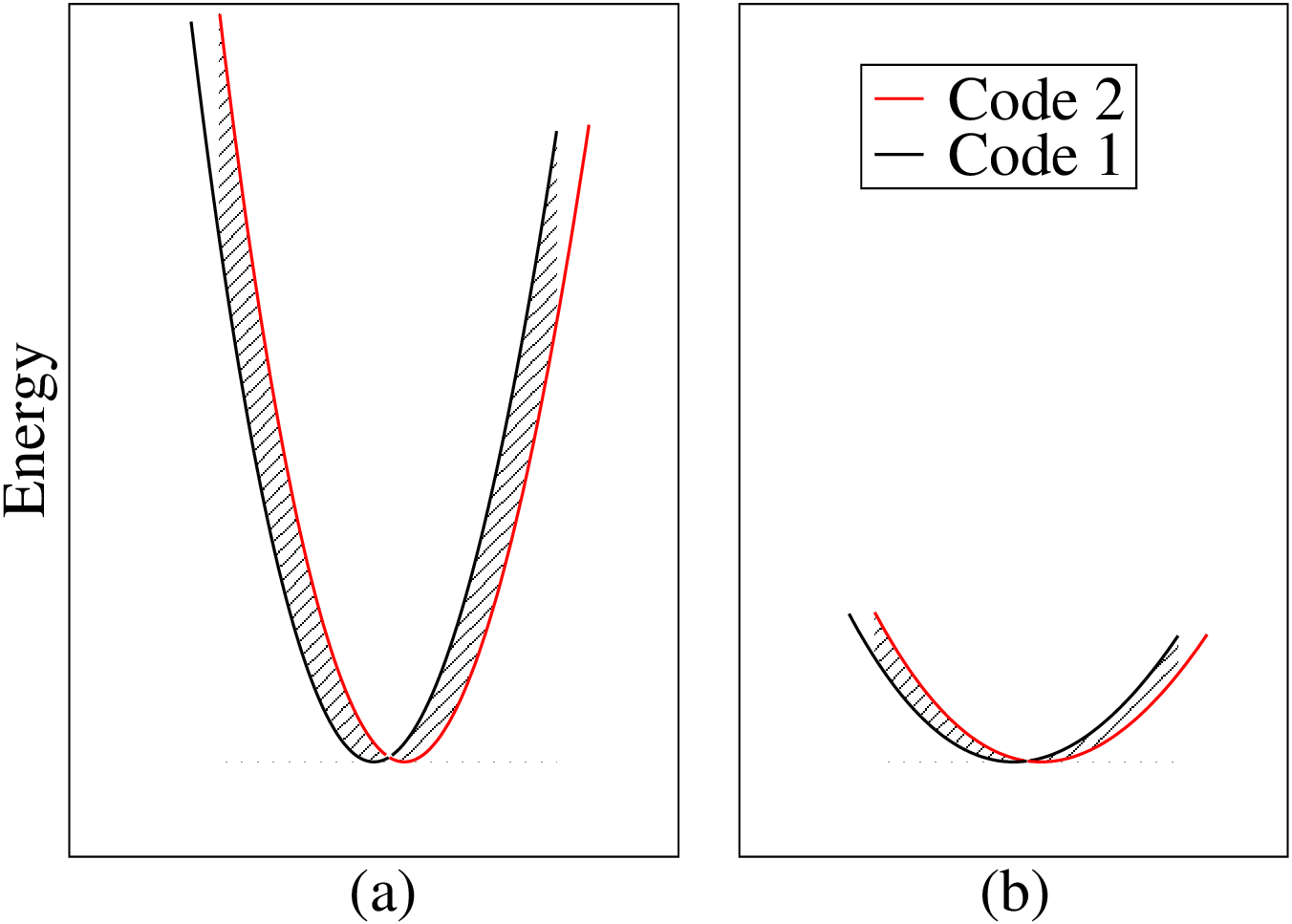}}}} \\
\caption{Comparison of the EOS obtained by two codes (a) for a high value of the bulk modulus (b) for a small value of the bulk modulus}
\label{fig3}
\end{figure}
\end{center}

The reason why some elements are very sensitive to the accuracy of $V_0$, B and B' (compared to the AE values) comes from a great dispersion of the $V_0$ and B values over the elements. $V_0$ indeed ranges from 7.2 $Bohrs^{ \rm 3}$ for Boron to 117.7 $Bohrs^{ \rm 3}$ for Cs, whereas B ranges from 0.57 GPa for Ar to 401 GPa for Os, nearly three order of magnitude! This means that a very small deviation of $V_0$ from the reference AE value will give a very large value of the $\Delta$ factor for a high B value (fig.\ref{fig3}-a), whereas a large deviation of $V_0$ from the reference AE value will give again a very good value of the $\Delta$ factor for a low B value (fig.\ref{fig3}-b).
\\
 For instance, a $V_0$ deviation of 0.76{\%} for Cs leads to $\Delta_{Cs}=0.39 meV$ whereas a $V_0$ deviation of 0.76{\%} for Os leads to $\Delta_{Os}=9.14 meV$ (ABINIT calculation with the PAW 0.9 atomic data). The same effect happens between elements that have close bulk moduli but large differences for the equilibrium volumes.\\
 To overcome this difficulty, we propose to define a $\Delta_1$ factor, which is the same of the $\Delta$ factor except it is "renormalized" to reference values of  $V_0$ and B for all the elements: indeed the Birch-Murnaghan energy is directly proportionnal to $V_0$ and B (see Appendix), and so, to first order the $\Delta$ factor. For each element, we therefore define:\\

\begin{equation}
\Delta_{1}= \frac{V_{ref}B_{ref}}{V_{AE}B_{AE}}\, \Delta 
\end{equation}

 The $\Delta_1$ factor is a re-scaled value of $\Delta$ to a reference material charaterized by an equilibrium volume $V_{ref}$ and a bulk modulus $B_{ref}$. This allows a comparison of the $\Delta_1$ factor of all the elements, as it is normalized to the same reference. We have chosen the values of  $V_{ref}=30 Bohrs^{ \rm 3}$ and $B_{ref}=100 GPa$ as they correspond approximatively to the mean values of $V_0$ and B over the 71 elements tested.\\
 We have then calculated the $\Delta$ and $\Delta_1$ factor for four energy cut-offs (12 Ha, 15 Ha, 20 Ha, 40 Ha) and the atomic data sets available with the ABINIT code: the GPAW PAW 0.9 package, the GBRV-v1 package \cite{Note1}, and our new package (named JTH). A fourth package is under building by the PWPAW group \cite{web4} but the work is in progress and we have not used this package in this paper. The results are presented in Table 1 and 2.
\begin{table}[h]
\begin{tabular}{llllccccc}
\\\hline
 $\Delta$ (meV)& 12 Ha & 15 Ha & 20 Ha & 40 Ha \\
\hline
 JTH (71 elements)   &  2.461  &  0.817 & 0.363 & 0.453 \\
 PAW 0.9 (68 elements) & 4.845   & 2.289 & 1.559 & 1.552 \\
 GBRV-v1 (63 elements)  & 4.486  &  2.617 & 2.420 & 2.345
\\\hline
\end{tabular}
\caption{Comparison of the $\Delta$ values as a function of cut-off energy for three PAW atomic data sets, as calculated with the ABINIT code.}
\label{tab1}
\end{table}

\begin{table}[h]
\begin{tabular}{llllccccc}
\\\hline
 $\Delta_1$ (meV)& 12 Ha & 15 Ha & 20 Ha & 40 Ha \\
\hline
 JTH (71 elements)   &  7.671  &  2.187 & 0.888 & 0.970 \\
 PAW 0.9 (68 elements) & 12.117   & 5.267 & 3.092 & 2.828 \\
 GBRV-v1 (63 elements)  & 8.243  &  5.698 & 5.363 & 5.155
\\\hline
\end{tabular}
\caption{Comparison of the $\Delta_1$ values as a function of cut-off energy for three PAW atomic data sets, as calculated with the ABINIT code.}
\label{tab2}
\end{table}

To give a comparison point, when using the values given in \cite{Lejaeghere} for the VASP package (71 elements)\cite{KresseJoubert}, we obtain $\Delta=1.920 meV$ and $\Delta_1=3.786 meV$ (As indicated in \cite{Lejaeghere}, the energy cutoff is 15 Ha for most elements and 22 Ha for He, B, C, N, O, F and Ne).\\
For the three packages (JTH, PAW 0.9, GBRV-v1), $\Delta$ and $\Delta_1$ are well converged for a 20 Ha energy cutoff. For all the energy cutoffs, the JTH package gives smaller $\Delta$ and $\Delta_1$ values than the two other ones. For $\Delta$, the difference is around 1.2 meV whereas for $\Delta_1$ the difference is arround 2 meV for converged energy cutoffs.

\begin{center}
\begin{figure}[h]
{\resizebox{13.0cm}{!}
{\rotatebox{0}{\includegraphics{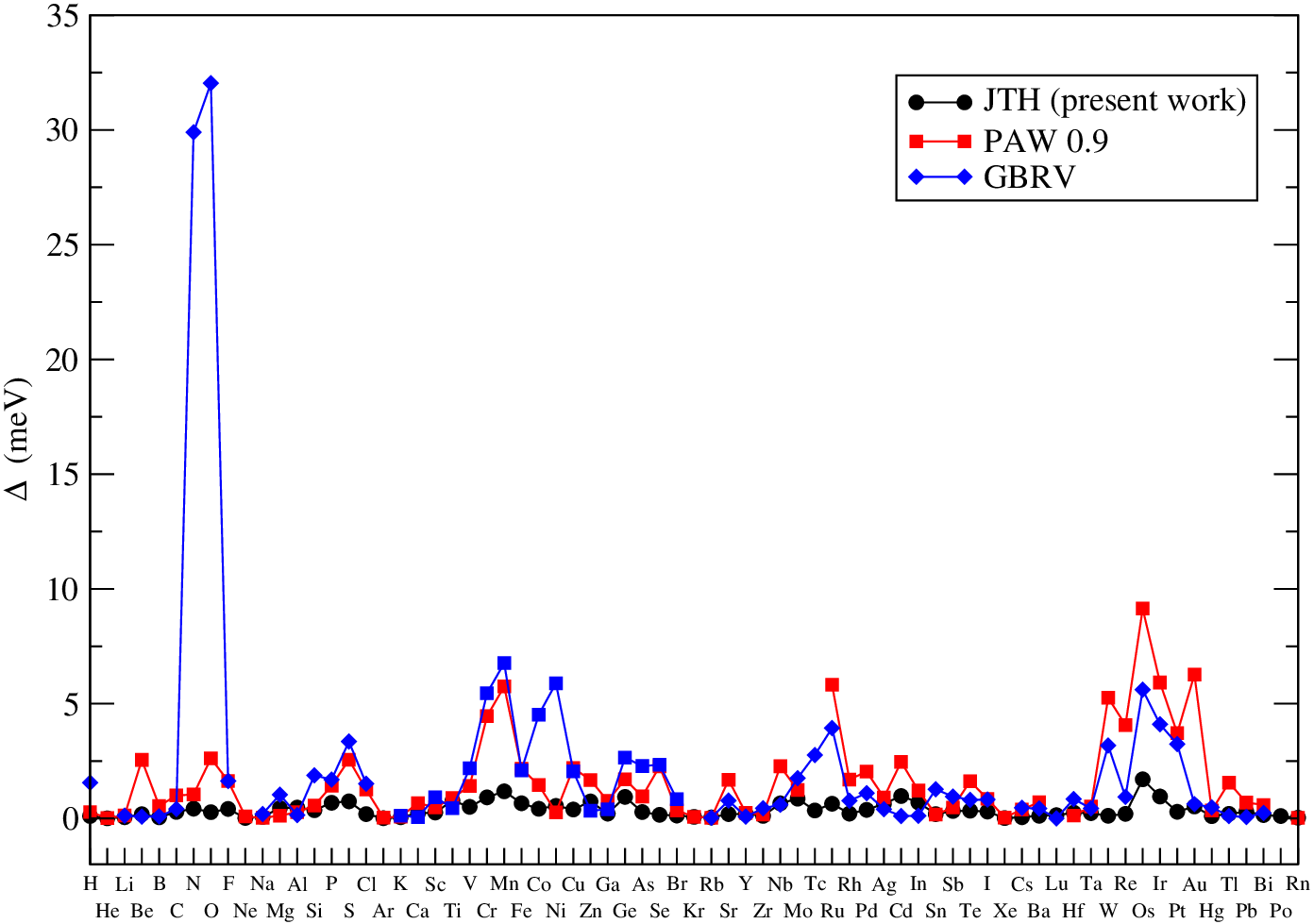}}}} \\
\caption{$\Delta$ value as a function of element for a 20 Ha energy cut-off: comparison between JTH, PAW 0.9 and GBRV}
\label{fig4}
\end{figure}
\end{center}
If we look in detail at the $\Delta$ factor (fig.\ref{fig4}), we can see that for the GBRV-v1 package, only two elements (N, O) are above 10 meV. Without these two elements, $\Delta$= 1.484 meV, which is a very good result when compared to the other packages\cite{Note} .  
\begin{center}
\begin{figure}[h]
{\resizebox{13.0cm}{!}
{\rotatebox{0}{\includegraphics{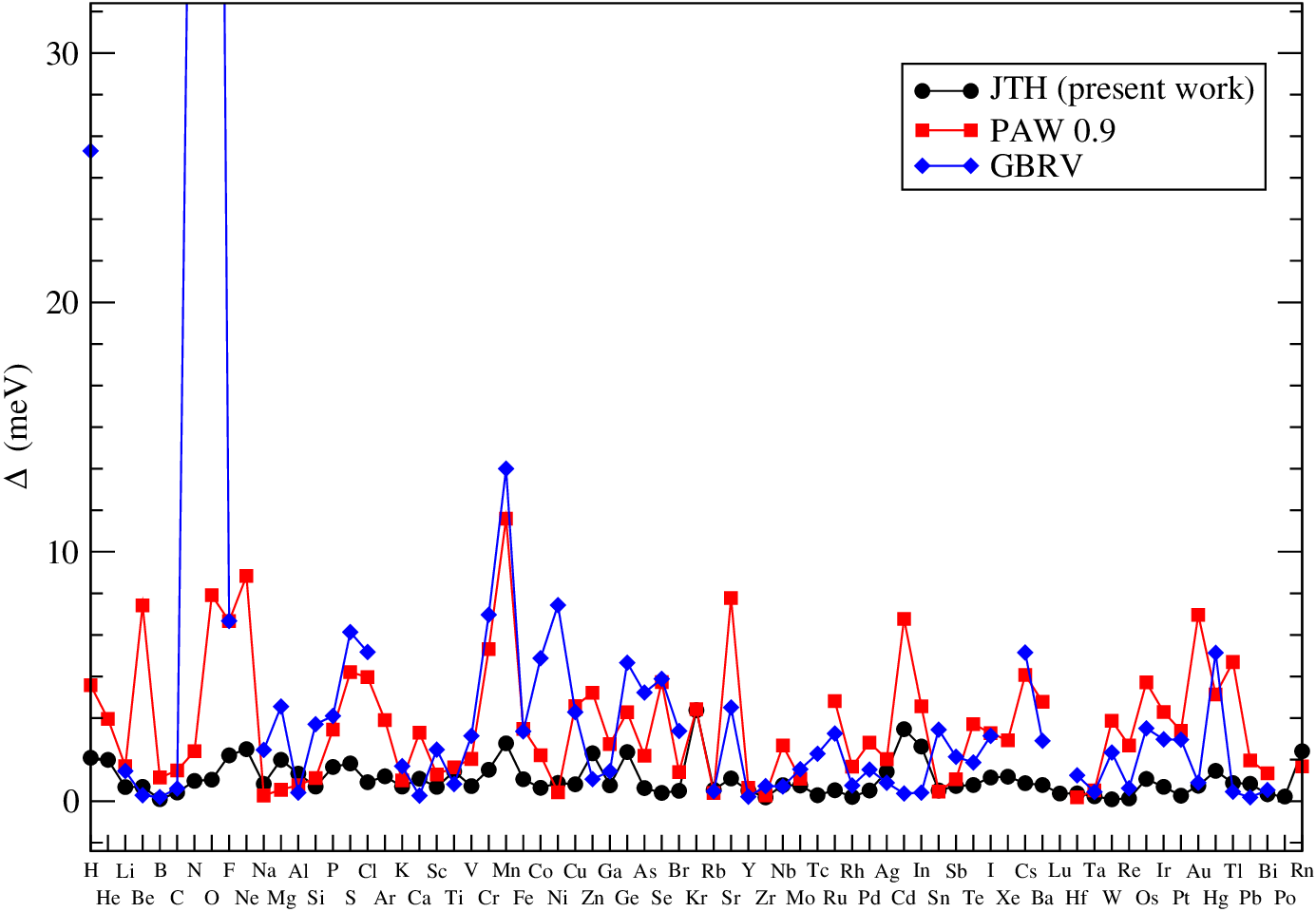}}}} \\
\caption{$\Delta_1$ value as a function of element for a 20 Ha energy cut-off: comparison between JTH, PAW 0.9 and GBRV}
\label{fig5}
\end{figure}
\end{center}
For which concerns $\Delta_1$,(fig.\ref{fig5}), the same trends are found: without H, N, O ($\Delta_1 > 20 meV$), $\Delta_1$= 2.944 meV for the GBRV-v1 package.  The $\Delta_1$ factor, treating low and high equilibrium volume and bulk modulus element on an equal footing, allows to focus on questionable elements, the $\Delta_1$ factor of which is very high compared to the mean $\Delta_1$.
It is also noticable that the JTH package has quite low values of $\Delta$ and $\Delta_1$ for a 15 Ha energy cutoff, which is essential in the frame of high-throughput calculations. 

\section{Conclusions}

Thanks to the flexibility of the ATOMPAW generator, we have been able to generate a 71 elements PAW dataset table. This JTH table has been validated against AE calculations thanks to the $\Delta$ factor and to the modified $\Delta_1$ factor we have defined in this paper. The JTH table has good accuracy and efficiency compared to other packages makes it a good candidate for high-throughput calculations. This new table is provided as XML files, that makes it easily readable by all the PAW codes. It is distributed on the ABINIT web site \cite{web3}.

\section{Acknowledgments}

The authors thank Bernard Amadon and Kevin Garrity for helpful discussions concerning the generation of PAW atomic data. This work was partly performed using HPC resources from the French Research and Technology Computing Center (CCRT).  
The contributions by NH to this effort
was supported by NSF Grant No. DMR-1105485.

\appendix
\section{}
The Birch-Murnaghan energy

\begin{equation}
E(V)-E_{0}=\frac{9V_{0}B}{16} \{[(\frac{V_{0}}{V})^{2/3}-1]^{3}B'+[(\frac{V_{0}}{V})^{2/3}-1]^{2}[6-4(\frac{V_{0}}{V})^{2/3}]\}
\end{equation}
is proportional to B, so that obviously, the $\Delta$ factor also: if one element A has a bulk modulus $B_{A}=\alpha B_{C}$, where C is another element, $\Delta_{B}\simeq\alpha \Delta_{C}$, with the hypothesis that $B_{A}^{code1}=\alpha B_{C}^{code1}$ and $B_{A}^{code2}\simeq\alpha B_{C}^{code2}$.\\
This is the same thing for the dependance against $V_{0}$, although it is more tedious to establish:\\
Let us consider 2 elements A and C and suppose that $V_{0}^{code1}(A)=\alpha V_{0}^{code1}(C)$ and $V_{0}^{code2}(A)=\alpha V_{0}^{code2}(C)$ for simplicity (with the same B and B' for the two elements). The segment in which $\Delta$ is computed is defined by $V_{i}=0.94 V_{S}$ and $V_{f}=1.06 V_{S}$ where $V_{S}$ is the center of the segment. $V_{S}$ is close to $V_{0}^{code1}$ and $V_{0}^{code2}$, and for simplicity, we suppose that $V_{S}(A)\simeq\alpha V_{S}(C)$.\\
As shown in the Appendix of \cite{Lejaeghere},
\begin{equation}
\Delta= \sqrt{\frac{F(V_{f})-F(V_{i})}{V_{f}-V_{i}} }
\end{equation}
where 
\begin{equation}
F(V)= \int_{V_{i}}^{V_{f}}{(E_{code1}(V)-E_{code2}(V))^{\rm 2} dV}=\sum_{\rm n=-2}^{\rm 4} x_{n} V^{-(2n+1)/3}
\end{equation}
using the definition of $x_{n}$ given in \cite{Lejaeghere}.\\
It can then been shown that each of the seven terms contributing to F is proportional to $\alpha^{3}$. For instance, for n=4, the contribution to $F_{A}(V_{f})$ is:
\begin{eqnarray}
x_{4}^{A}(V_{f}^{A})^{-3}&=& -\frac{1}{3}(\frac{9(V_{0}^{code1}(A))^{3}B_{A}^{code1}}{16} (B'_{code1}(A)-4)-\frac{9(V_{0}^{code2}(A))^{3}B_{A}^{code2}}{16} \nonumber \\ 
 &&(B'_{code2}(A)-4)))^{2}(1.06V_{S}(A))^{-3} \nonumber \\
&\simeq&-\frac{\alpha^{6}}{3}(\frac{9(V_{0}^{code1}(C))^{3}B_{C}^{code1}}{16} (B'_{code1}(C)-4)-\frac{9(V_{0}^{code2}(C))^{3}B_{C}^{code2}}{16} \nonumber \\ 
 &&(B'_{code2}(C)-4)))^{2}(1.06{\alpha}V_{S}(C))^{-3} \nonumber \\
&\simeq& {\alpha}^{3} x_{4}^{C}(V_{f}^{C})^{-3}
\end{eqnarray}

We have therefore $F_{A}(V_{f})\simeq{\alpha}^{3}F_{C}(V_{f})$, $F_{A}(V_{i})\simeq{\alpha}^{3}F_{C}(V_{i})$ and $V_{f}^{A}-V_{i}^{A}\simeq\alpha(V_{f}^{C}-V_{i}^{C})$.\\
So, at the end:
\begin{equation}
\Delta(A)\simeq \alpha\Delta(C)
\end{equation}












\end{document}